\documentclass[a4]{iopart}
\usepackage{epsfig,graphicx,float}
\usepackage{amssymb}

\newcommand{\be}{\begin{equation}} 
\newcommand{\ee}{\end{equation}}
\newcommand{\bc}{\begin{center}}
\newcommand{\ec}{\end{center}}

\begin{document}

\title{Detailed analysis of quantum phase transitions within the $u(2)$ algebra}
\author{L.Fortunato$^{1,\footnote{On leave for the ECT*, Strada delle tabarelle 286, I-38050 Villazzano (TN), Italy}}$ and L.Sartori$^2$}
\address{1- Dipartimento di Fisica ``G.Galilei'' and INFN, via Marzolo 8, 
I-35131 Padova, Italy.}
\address{2- CISAS "G. Colombo", via Venezia 15, 35131, Padua, Italy}

\begin{abstract}
We analyze in detail the quantum phase transitions that arise in models based on the $u(2)$
algebraic description for bosonic systems with two types of scalar bosons. 
First we discuss the quantum phase transition that occurs in hamiltonians that admix the two dynamical
symmetry chains $u(2)\supset u(1)$ and $u(2)\supset so(2)$ by diagonalizing the problem exactly in the $u(1)$ basis. 
Then we apply the coherent state formalism to determine the energy functional. Finally we show that a quantum
phase transition of a different nature, but displaying similar characteristics, may arise also within a single chain just by including higher order terms in the hamiltonian.
\end{abstract}
\maketitle

\section{Introduction}
Quantum phase transitions are a subject of current interest in various branches of physics that range from molecular to nuclear and to hadronic physics \cite{Ro,Cej, Cej2, Cap}. 
Often one describes a certain set of measurable properties of a quantum system with phases that 
have a genuine quantum origin and, within some simplified algebraic model, tries to associate these regimes with the 
manifestation of a given (dynamical) symmetry. These symmetries are connected with Lie algebras and, 
by exploiting their mathematical properties, one is usually able to make detailed predictions for important observables 
and classify large sets of data using recurring patterns and schemes that proceed from the formal 
algebraic structures. 
These methods have been heavily employed in nuclear and molecular physics, for example in the Interacting Boson Model (IBM) and Vibron model\cite{IL,FvI}, that make use of the $u(6)$ and $u(4)$ algebras respectively.
However, due to the high dimension of the algebras involved, it would be easier to discuss the simplest possible case to
highlight that several important features, discussed in connection with the IBM or other algebraic models, are extremely common (and very likely universal) as they appear even in the paradigmatic case of $u(2)$. This algebra is often found in connection with the schematic Lipkin model, that is often taken as an example of solvable many-body system.
Two subalgebras may occur within the $u(2)$ algebra: one, $u(1)$ is usually associated with a vibrational spectrum and with a 'spherical' phase, while the other, $o(2)$, is associated with a rotational spectrum and with a 'deformed' phase. In the following we will study the $u(1)$-$o(2)$
phase transition (the only possible within this scheme), by solving a transitional hamiltonian in the $u(2) \supset u(1)$ 
basis. We will give energy spectra along the transition and study the phase transition in terms of the mixing of components of the original eigenvectors. The analysis of the composition of the eigenvectors reveals a remarkable 
persistence of the vibrational character during most of the phase transition up to the critical point. 

In parallel to the analytic calculations that made within the Lie algebraic scheme, 
coherent states have been used to connect the abstract algebraic formalism to a geometric description by means of 
a mapping of second quantization operators in terms of differential operators. This procedure allows
to transform the algebraic hamiltonian into a Schr\"odinger-like differential equation that can be used
to calculate a potential energy surface (the expectation value of the hamiltonian in the ground state). 
The position of the minima of this functional determines in turn several important properties of the initial system 
and furnishes a way to study and interpret the phases and the possible phase transitions as well as their dependence on certain parameters. 
The coherent state has been heavily employed in the IBM, in the Vibron model and in several other models. This 
method is crucial to define the critical point and the character of the phase transition.

Moreover, we will use both the exact diagonalization and the coherent state formalism to show how a quantum phase transition between two different behaviours can be obtained within a single subalgebra chain, just by allowing a combination of linear and quadratic Casimir operators of $u(1)$.

The $u(2)$ algebra has been investigated in the thesis of O.van Roosmalen \cite{vR}, in several books about algebraic models \cite{FvI} and in a recent paper by Cejnar and Iachello\cite{Cej}, that treats the $u(n-1)$-$o(n)$ quantum phase transition in general. Our paper provides a detailed discussion of the simplest specific case of $u(2)$. 

In Ref. \cite{Vid} a scalar two-level boson model is set up to simulate the consistent Q formalism (CQF) for the Interacting Boson Model getting rid of the complications of the $d_\mu$ quadrupole boson operators. This allows the model to be solved for a much larger number of bosons, that is a severe limitation in the usual CQF. We adopt a different perspective in our study as we don't define the hamiltonian to allow a comparison with IBM.

In Sect. (\ref{II}) we repeat a few basic definitions and we study the eigenvalues and eigenstates along the phase transition, in Sect. (\ref{III}) we calculate matrix elements of the relevant Casimir operators and we write the potential energy functional associated to the phase transition, while in Sect. (\ref{IV}) we study the 
consequences of the introduction of higher order Casimir operators into the hamiltonian.

\section{$u(1)-so(2)$ transition}
\label{II}
The book of A.Frank and P.van Isacker \cite{FvI} provides a thorough pedagogical discussion of the $u(2)$ 
dynamical symmetry and in particular of the two symmetry limits
in the solution of the s-t-boson hamiltonian. The limits correspond to the subalgebras $u(1)$ and $so(2)$ and are 
associated with an anharmonic oscillator (or spherical phase) and with a rotor (or deformed phase) respectively.
One can think of this algebra as the simplest possible algebra with applications to a complex system, made up
of different interacting scalar bosons. We remind here only the essential formulas and we refer the reader to Ref. \cite{FvI} 
for further details. Notice that there are several differences in the definitions with respect to Ref. \cite{vR} that make the comparison a little laborious. We will adhere to the conventions in \cite{FvI}.

The basic building blocks are creation and annihilation operators for the two types of scalar bosons, called $s$ and $t$.
Their commutation relations are 
\be
[s,s^\dag]=[t,t^\dag]=1 \;,
\ee
all the others being zero.
To distinguish the two kinds of bosons one can arbitrarily define a parity operator, $\hat P$, as follows \cite{FvI}:
\be
\hat P s^\dag \hat P^{-1}= s^\dag \qquad \hat P t^\dag \hat P^{-1}= -t^\dag \;.
\ee

The following four bilinear operators,
\be
\hat n_s=s^\dag s \qquad s^\dag t \qquad t^\dag s \qquad \hat n_t=t^\dag t
\label{bo}
\ee
close under commutation according to the rules that define the $u(2)$ Lie algebra.
Two of them give directly the number operators for the boson of species $s$ and $t$, while 
their sum gives the total number operator, i.e.:
\be
\hat N =s^\dag s + t^\dag t \;.
\ee

\begin{figure}[!h]
\includegraphics[clip=,width=0.8\textwidth]{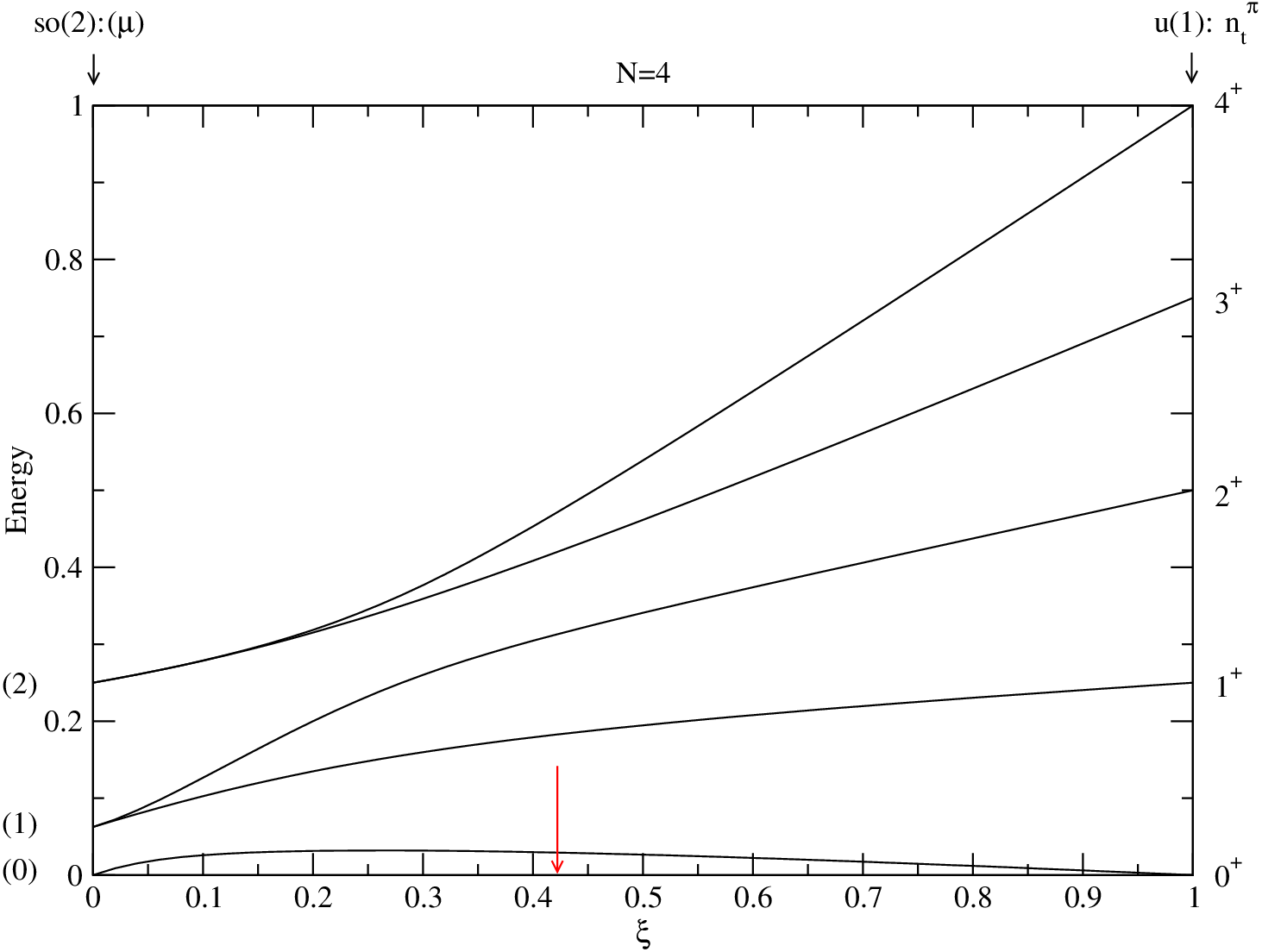}
\caption{Eigenstates of the transitional hamiltonian as a function of $\xi$, for $N=4$. The two dynamical symmetries (black arrows) are recovered at the edges, where the labeling is also given according to \cite{FvI}. The red arrow marks the critical point (see section 3).}
\label{N=4}
\end{figure}

\begin{figure}[!h]
\includegraphics[clip=,width=0.49\textwidth]{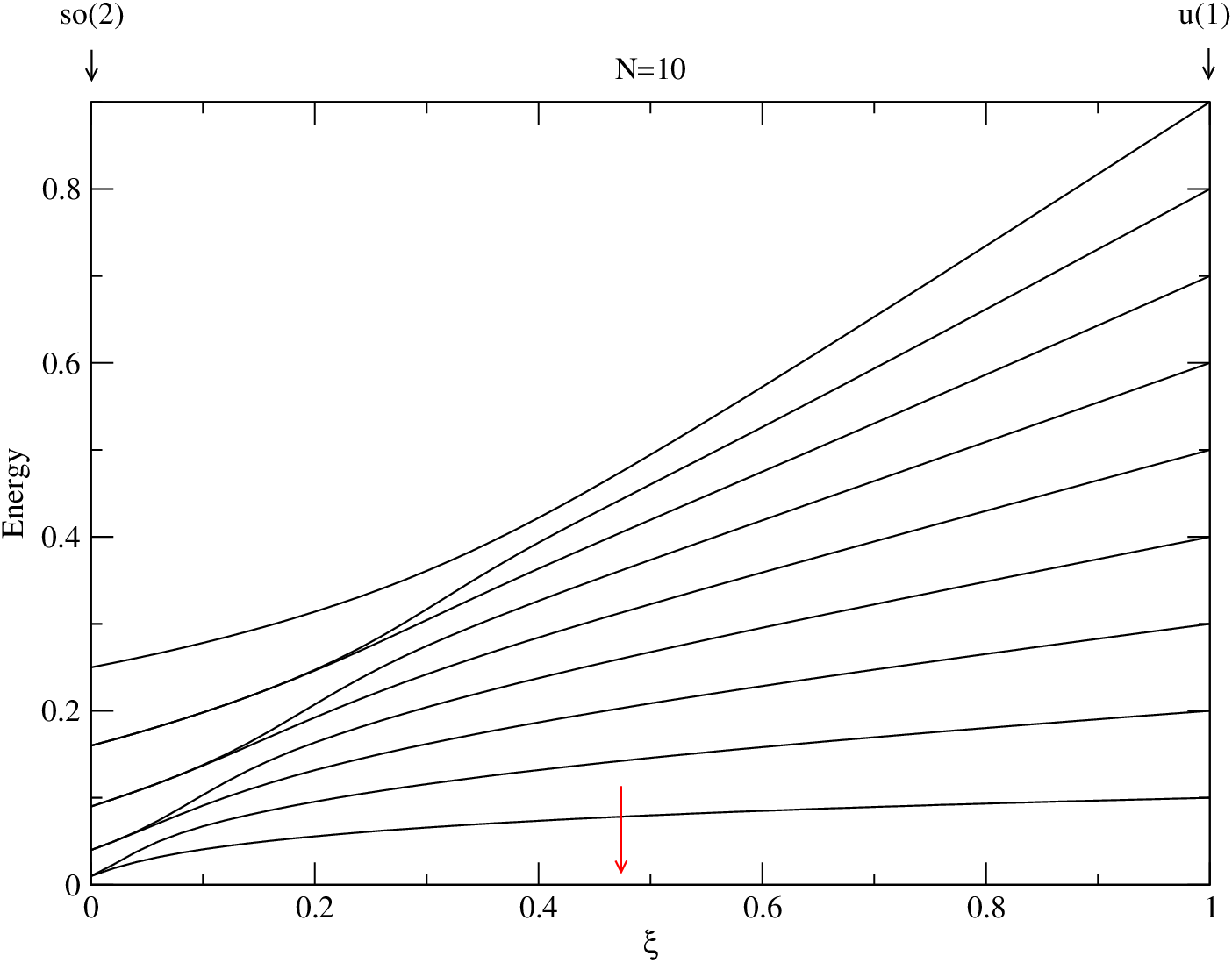}
\includegraphics[clip=,width=0.49\textwidth]{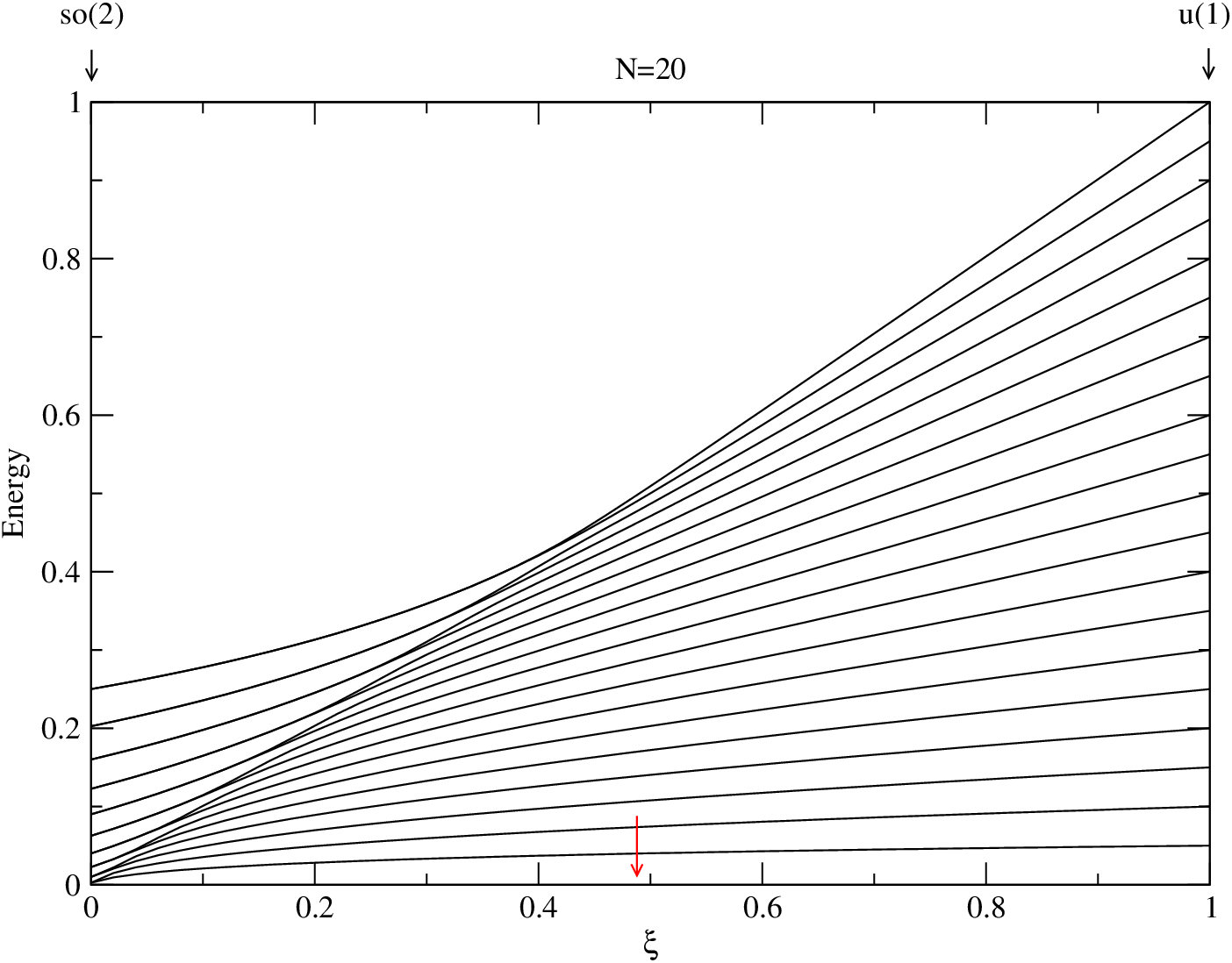}
\caption{Same as Fig. \ref{N=4}, but for N=10 and N=20 bosons. The arrows mark the occurrence of the critical points, according to Eq.(\ref{cp}) with $\alpha=0$.}
\label{N=10}
\end{figure}

The bilinear operators in (\ref{bo}) can also be rearranged into several physically 
meaningful operators, such as, for example, the components of a angular momentum (see \cite{FvI}).
In what follows we shall only need the square of the third component of this angular momentum, namely:
\be
\hat J_z^2 = \frac{1}{4}(\hat N - s^\dag s^\dag tt - t^\dag t^\dag ss +2s^\dag t^\dag st ) \;.
\label{cu2}
\ee
One can then define the most general one- and two-body hamiltonian in the s-t space as
\be
\hat H=E_0+\epsilon\hat n_t+\alpha \hat n_t^2+\beta \hat J_z^2 \;.
\label{Hami}
\ee
In general, apart from the constant term, the linear and quadratic terms in $\hat n_t$ satisfy the 
$u(2)$$\supset$$u(1)$ chain, while the last term satisfies the $u(2)$$\supset$ $so(2)$ chain.
These two dynamical symmetry chains are the only possible chains within $u(2)$.
It is appropriate to rescale each of the terms in the above equation by the correct power of $N$ to ensure that the 
critical point will remain independent of $N$ in the large $N$ limit (at the leading order). This means that 
the one-body term (in this case $\hat n_t$) must be divided by $N$ and each two-body term by $N^2$ as follows:
\be
\hat H=E_0+\frac{\epsilon\hat n_t}{N}+\frac{\alpha \hat n_t^2}{N^2}+\frac{\beta \hat J_z^2}{N^2} \;.
\label{HamiR}
\ee
We begin our analysis by considering a generic hamiltonian as a linear combination of terms with definite symmetry
\be
\hat H= \xi \hat H_{u(1)} + (1-\xi) \hat H_{so(2)} \;,
\label{H1H2}
\ee
with $\hat H_{u(1)}=\hat n_t/N$ and $\hat H_{so(2)}= \hat J_z^2/N^2$.
The first step consists in studying the behaviour of the spectrum as a function of $\xi$ 
for a given boson number. We do this here diagonalizing exactly the hamiltonian
(\ref{H1H2}) in the $u(1)$ basis. Notice that the only non-diagonal matrix elements in this
basis are the ones of $\hat J_z^2$, that are given explicitly in Ref. \cite{FvI}.  For the sake of simplicity 
the spherical part has been restricted to the purely linear term, $\hat n_t$, in order to obtain a 
harmonic oscillator when $\xi=1$.  In other words, Eq. (\ref{H1H2}) has been obtained from Eq. (\ref{HamiR}) 
setting $\alpha=0$, $\epsilon=\xi$ and $\beta=1-\xi$. The results of the numerical diagonalization are given in
Fig. \ref{N=4} for $N=4$ bosons and Fig. \ref{N=10} for $N=10$ and $N=20$.
The first figure allows a direct comparison with the figures contained in Ref. \cite{FvI}.
One, of course, recovers the exact analytic solutions when $\xi$ takes the two values at the 
limits of the interval $[0,1]$, corresponding to the two dynamical symmetry chains $u(2) \supset so(2)$ and 
$u(2)\supset u(1)$. 
At these points the quantum numbers coming from the two dynamical symmetries discussed above are 
strictly valid and are given on the two sides of the figure: $\mu$ labels pure $so(2)$ states and $n_t^\pi$ labels
pure $u(1)$ states.

The energy level pattern dependence on the transition parameter for boson numbers 10 and 20 is displayed in 
Fig. \ref{N=10}.
 
This allows to compare the density of states of the two regimes (quadratic on the left and linear on the right side respectively)
and to see that the additional degeneration of the $so(2)$ shifts to the right for excited states.
Notice that these pictures are very similar to the corresponding diagrams (for the U(3) case) of Ref. \cite{Cej}.

\begin{figure}[!h]
\includegraphics[clip=,width=1.\textwidth]{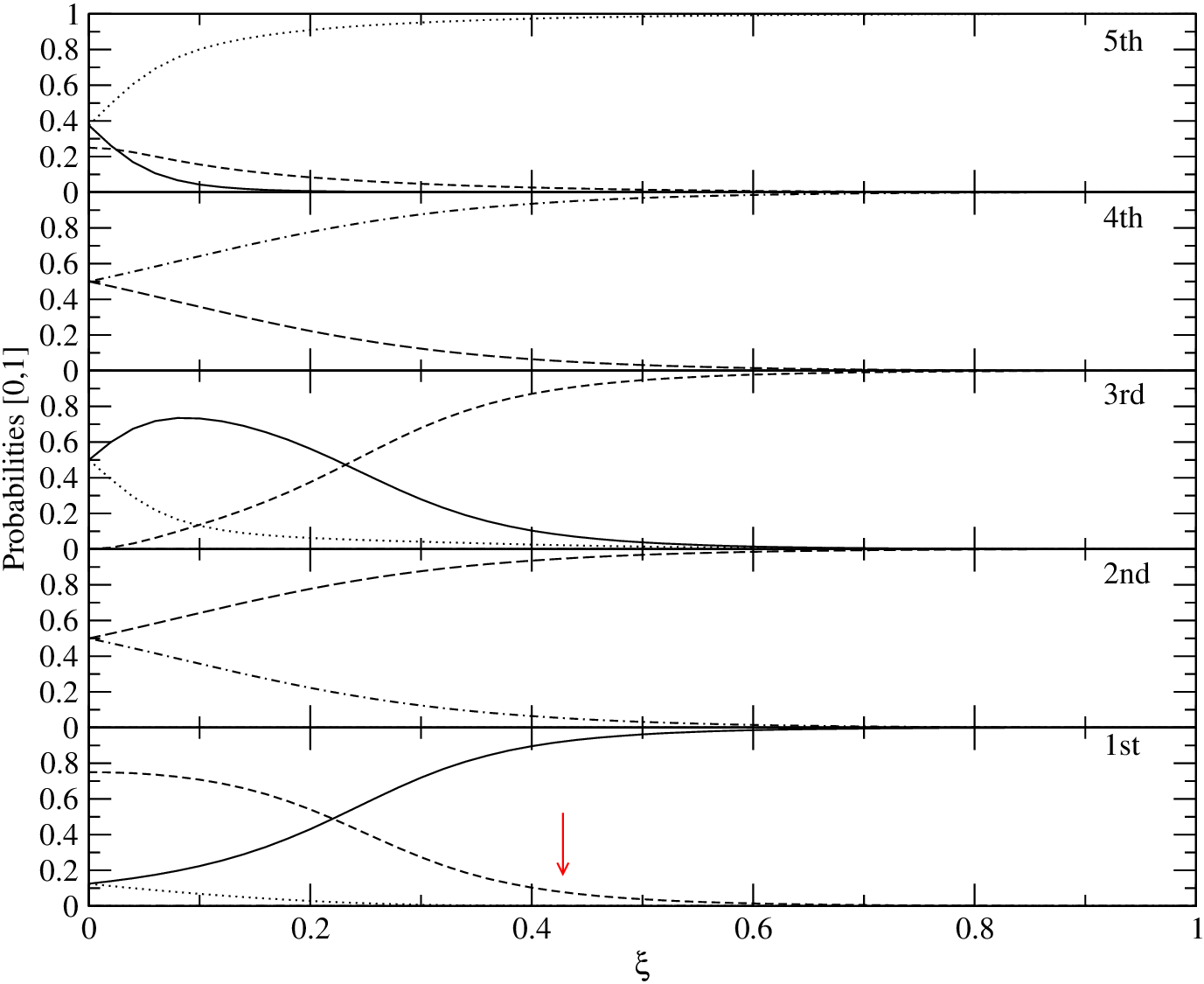}
\caption{Components, given as probabilities in the interval $[0,1]$, of the transitional eigenstates 
as a function of $\xi$ in the $u(1)$ basis for the $N=4$ case. The pure $u(1)$ eigenstates are indicated with full, long-dashed, 
short-dashed, dot-dashed and dotted lines for $n_t=0,\cdots, 4$ respectively. Of course only states with 
the same parity appear as components of a given state.}
\label{compo}
\end{figure}

The diagonalization of the hamiltonian matrix given in Eq. (\ref{H1H2}) provides also the eigenvectors. 
We have plotted in 
fig. \ref{compo} the squared amplitudes of the five eigenvectors of the case $N=4$ with respect to the 
components of the $u(1)$-basis. At the limit $\xi=1$ one clearly sees that these eigenvectors are pure in the 
$u(1)$ basis, but, moving far from the limit, they acquire components from other states (with same parity).
For example the ground state remains quite pure above the value $\xi\sim 0.42$ that corresponds to the 
critical point and it becomes completely mixed only in the vicinity of the other dynamical symmetry limit.
Similar considerations can be drawn also for the excited states, although with the due differences: for example
a larger mixing is seen to take place for smaller values of $\xi$ in the third and fifth states.
In the present case the second and fourth state, being the only ones with 'negative' parity mix in a well-ordered 
fashion, just reversing the role of the highest and lowest components.

\begin{figure}[!h]
\includegraphics[clip=,width=1.\textwidth]{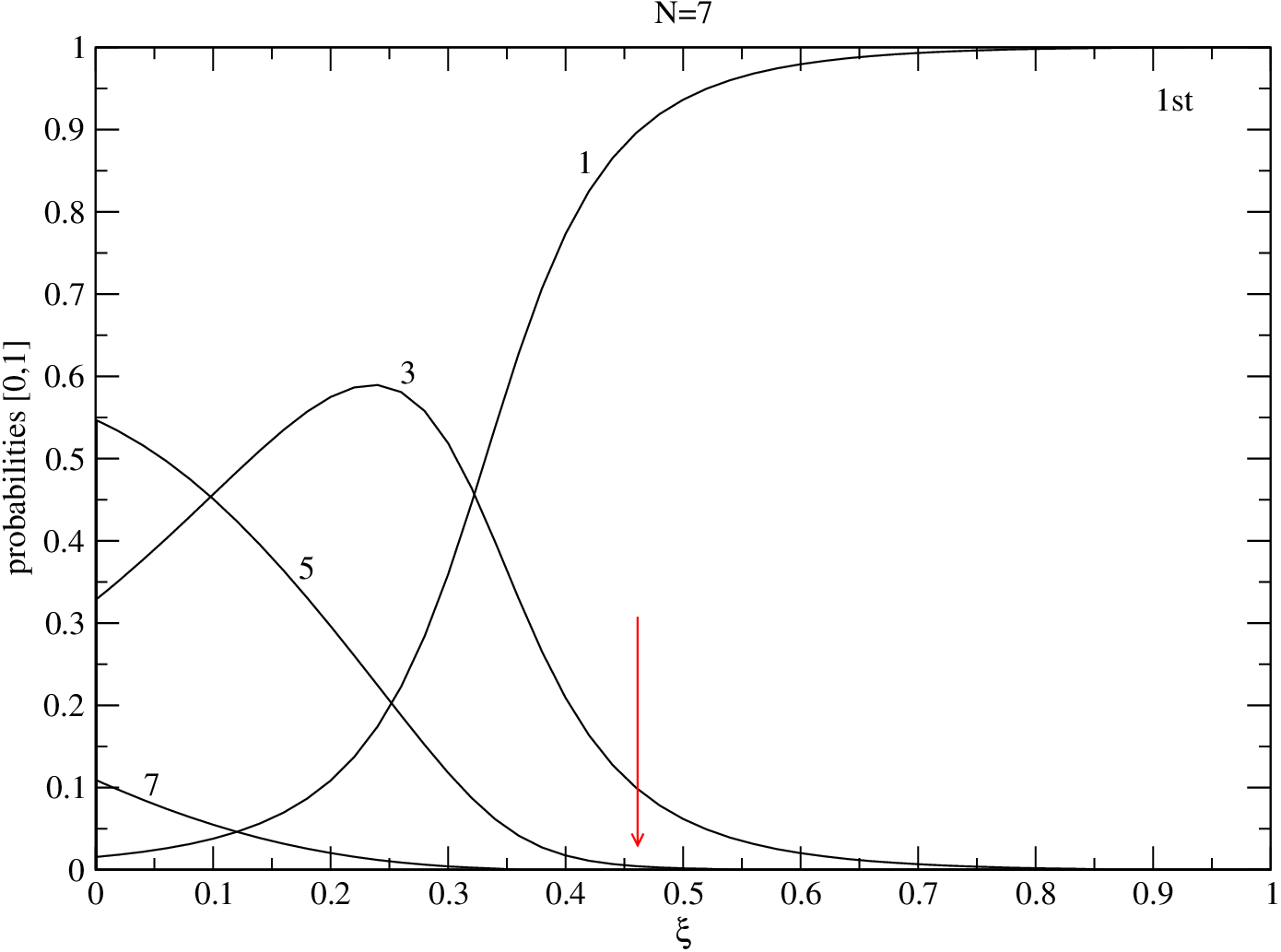}
\caption{Components, given as mixing probabilities, of the lowest eigenstate 
as a function of $\xi$ in the $u(1)$ basis for the $N=7$ case. 
Numbers from $1$ to $7$ correspond to the quantum number $n_t$ of the limiting $\xi=1$ case. }
\label{compo7}
\end{figure}

If one repeats the calculation of the squared amplitudes for larger boson numbers $N$, it is seen that,
 the critical point approaches the limit $\xi =1/2$, and the eigenstates keep as a largest component 
the one that would correspond to the spherical phase ($u(1)$ limit) above that point. This fact was already noticed by Iachello and several other authors for different cases. 
As an example we have plotted in fig. \ref{compo7} the same as the lowest panel of fig. \ref{compo}, but for 
$N=7$. In this case the critical point sits at $\xi_c=6/13$, but the eigenvectors keep as a leading component the 
$n_t=0$ state of the limiting $\xi=1$ spherical case down to about $\xi \sim 0.35$. 
This fact has an important consequence: although the spectrum (the energy levels) displays large modifications already 
around the critical point, the eigenvector have a little resilience to abandon the spherical configuration.
This means that, proceeding from the spherical to the deformed case, all those spectroscopic properties, that depend heavily on the details of the wave functions (for example electromagnetic transitions, etc.), will tend to remain fairly similar to their 'spherical' counterpart, at least in a small interval around the critical point.

\section{Coherent states calculations}
\label{III}
The connection with a geometrical picture is obtained by resorting to the formalism of vector coherent states.
They can be defined, in the present work, as
\be
\mid z,N \rangle = \frac{1}{\sqrt{(|z_1|^2+|z_2|^2)^N}}\frac{1}{\sqrt{N!}}(z_1s^\dag +z_2t^\dag )^N \mid 0 \rangle \;,
\ee
where $z_1$ and $z_2$ are complex numbers. These states are normalized $\forall N$.
Following a standard method, based on the connection of the commutator with differentiation, to calculate matrix elements of the operators that are relevant to the present discussion we obtain:
\be
\langle z ,N \mid \hat n_t \mid z,N \rangle = N\frac{|z_2|^2}{|z_1|^2+|z_2|^2}
\ee

\be
\langle z,N \mid \hat n_t^2 \mid z,N \rangle = N\frac{|z_2|^2}{|z_1|^2+|z_2|^2}+N(N-1)\frac{|z_2|^4}{(|z_1|^2+|z_2|^2)^2}
\ee

\be
\langle z,N \mid \hat J_z^2 \mid z,N \rangle = \frac{N}{4} + \frac{N(N-1)}{4}\frac{(z_1^*z_2-z_1z_2^*)^2}{(|z_1|^2+|z_2|^2)^2}  \;.
\ee

Among all the possible choices of $z_1$ and $z_2$, several would cancel the two-body term, as it occurs for instance when  $z_1$ and $z_2$ are taken as real numbers. One choice that allows us to retain this term is to take $z_1=1$ and $z_2=iy$ purely imaginary. Of course other choices are possible, provided that the matrix element don't become zero. Thus the energy functional associated with the hamiltonian (\ref{HamiR}) is given by:
\be
{\cal E}(x)=  \frac{\beta}{4N} + x^2 \Bigl(\epsilon+\frac{\alpha}{N}-\beta \frac{N-1}{N}\Bigr) + x^4 \frac{N-1}{N} (\alpha+\beta) \;,
\ee
where we have further introduced  $x^2=y^2/(1+y^2)$. We notice that the combination of $x^2$ and $x^4$ can be interpreted as a Landau potential, that leads to a second order phase transition. The critical point is obtained when the potential is purely quartic or, equivalently, by requiring the coefficient of the quadratic term to be zero. One obtains the following expression for the critical point:
\be
\xi_c=\frac{N-1-\alpha}{2N-1} 
\label{cp}
\ee
that has been used in the case $\alpha=0$ to mark all the red arrows in the preceding figures.
We point out here that, in the large boson limit ($N\rightarrow \infty$), the critical point tends to $1/2$. The presence of the quadratic Casimir of $u(1)$ in the hamiltonian can modify the critical point and the whole phase transition considerably.

\section{Higher order Casimir of $u(1)$}
\label{IV}
In section 4 we have considered only the linear Casimir operator
of $u(1)$ in the hamiltonian entering Eq. (\ref{HamiR}), that is $\alpha=0$. This was partly due to a need for simplicity, but also because the $\hat n_t^2$ term has a z-space realization that incorporates a $z^4$ term.
Conventional wisdom on quantum phase transitions and a large part of current literature on this topic associate 
(as we also have implied in the preceding sections) a phase with a dynamical symmetry chain. This is most often the case, because different symmetry chains contribute with different Casimir operators to the final energy functional. There 
might be cases in which, within a given symmetry chain, one could opt for including higher order terms, that, under
 certain choices for the parameters, can generate different phases, and perhaps give the same overall behaviour that was obtained by 
 mixing up different subalgebra chains. Albeit extremely simple, the algebra discussed here allows us to discuss this general statement with a crystal-clear example: we take only the subalgebra chain passing through the unitary subalgebra, namely 
$u(2)\supset u(1)$, with a combination of  both the linear and the quadratic Casimir operators (therefore there is no term depending on $\hat J_z^2$ contrarily to the previous section, or in other words $\beta=0$) that reads:
\be
\hat H_I=\epsilon \frac{\hat n_t}{N}+ \alpha \frac{\hat n_t^2}{N^2}
\ee
with spectrum  $E_I=\epsilon n_t/N+ \alpha n_t^2/N^2 $.
This is identical to the dynamical symmetry discussed in Ref. \cite{FvI}, sect. 1.4, save for the fact that we have dropped the constant term. 
One can again use the equations of the preceding section to determine the potential energy surface to be :
\be
{\cal E'}(x)=(\epsilon+\alpha) N x^2+\alpha N(N-1)x^4
\ee
that has a second order critical point when $\epsilon=-\alpha, \forall N$. It is clear that one has a 'spherical' minimum when $\epsilon$ is above that threshold and a 'deformed' minimum otherwise. Thus two different phases
can be generated within just one dynamical symmetry chain, without any need for mixing up with another symmetry, but just 
by including higher order terms. By higher order we mean powers of bilinear operators that, upon canonical ordering, might contain two-body terms, or higher order terms. In this case the deformed behaviour arises because of the presence of two-body terms coming from $\hat n_t^2$, rather than because of the mixing with two-body terms contained into $\hat J_z^2$.
This case could be called anharmonic vibrator, because of the quadratic term in the spectrum. 
The foregoing discussion allows us to conclude that a 'vibrator' is not necessarily associated with a 
'spherical' minimum although its underlying algebra are still $u(2)$ and $u(1)$, but indeed here we find a range of parameter values that allows for a 'deformed' minimum.
In summary: when $\alpha=0$ or when $\epsilon=0$ one only has a minimum in zero; when $\alpha>0$ one may have either a minimum in zero for $\epsilon > -\alpha$, the critical point as discussed above and a deformed minimum for $\epsilon < -\alpha$.

\section{Conclusions}
The study of the $u(2)$ algebra has been extended from the pure dynamical symmetries to the transitional region between them. 
The whole transitional path is covered by constructing a hamiltonian that allows to pass with continuity from $u(1)$ to $so(2)$ by varying a parameter.
Spectra and eigenstates are calculated by diagonalizing this hamiltonian in the $u(1)$ basis.
The ground state critical point is found by direct calculation of the potential energy surface within a coherent state formalism
that must be defined with a non-standard choice of phases.
%(one would have had a more standard situation by choosing $J_y$ as diagonal component of the angular momentum)

In spite of what one might expect from several studies on quantum phase transitions in higher dimensional algebras, in this low-dimensional case the critical point cannot be associated with the $e(1)$ symmetry. 
There are mainly two reasons for this fact. Firstly we are dealing here with a finite number of particles in an exact microscopic way, rather than solving a differential equation with an infinite square well potential in the $N\rightarrow \infty$ limit. Therefore we cannot expect to determine a dynamical symmetry at the critical point as it happens for example in five dimensions in the solution of the Bohr Hamiltonian in nuclear physics \cite{Iac1} or in two dimensions for the solution of the B\'es equation for collective nuclear pairing \cite{Cla}. Secondly the euclidean group in one dimension would have a very peculiar structure since $so(1)$ does not exists and rotations in one dimension essentially reduce to translation. 

Finally, a particular case including higher order operators of $u(1)$ has been discussed. This allows to show that, despite common belief, second order Casimir operators of $u(1)$ trigger the occurrence of a 'deformed' phase.
 This fact, that we exemplify in a schematic $u(2)$ model, to our knowledge has never been highlighted and its counterpart in more complicated algebras, such as those used in molecular and nuclear physics, might lead to a reconsideration of several algebraic models and introduce new facets in the description of various quantum systems.

\end{document}